\documentclass[12pt,preprint]{emulateapj}

\usepackage{rotating}
\usepackage{comment}
\usepackage{graphicx}
\usepackage{epstopdf}
\slugcomment{{\sc Accepted to ApJ:} March 12, 2014}

\shorttitle{Stochastic Modeling of the Fermi/LAT $\gamma$-ray Blazar Variability}
\shortauthors{Sobolewska et al.}

\begin{document}

\title{Stochastic Modeling of the Fermi/LAT $\gamma$-ray Blazar Variability}

\author{M. A. Sobolewska$^{1,2}$, A. Siemiginowska$^1$, B. C. Kelly$^3$, K. Nalewajko$^{4,5}$}
\affil{$^1$ Harvard-Smithsonian Center for Astrophysics, 60 Garden Street, Cambridge, MA 02138, USA\\
$^2$ Nicolaus Copernicus Astronomical Center, PAN, Bartycka 18, 00-716 Warsaw, Poland\\
$^3$ Department of Physics, Broida Hall, University of California, Santa Barbara, CA 93107, USA\\
$^4$ JILA, University of Colorado and National Institute of Standards and Technology, 440 UCB, Boulder, CO 80309, USA\\
$^5$ NASA Einstein Postdoctoral Fellow}
\email{malgosia@camk.edu.pl}

\begin{abstract}
We study the $\gamma$-ray variability of 13 blazars observed with the Fermi Large Area Telescope (LAT). These blazars have
the most complete light curves collected during the first 4 years of the Fermi sky survey. We model them with
the Ornstein-Uhlenbeck (OU) process or a mixture of the OU processes. The OU process has power spectral density (PSD)
proportional to $1/f^{\alpha}$ with $\alpha$ changing at a characteristic time scale, $\tau_{\rm 0}$, 
from 0 ($\tau \gg \tau_{\rm 0}$) to 2 ($\tau \ll \tau_{\rm 0}$). The PSD of the mixed OU process has two characteristic
time scales and an additional intermediate region with $0<\alpha<2$. We show that the OU model provides
a good description of the Fermi/LAT light curves of three blazars in our sample. For the first time we constrain
characteristic $\gamma$-ray time scale of variability in two BL Lac sources, 3C 66A and PKS 2155-304
($\tau_{\rm 0} \simeq 25$\,day and $\tau_{\rm 0} \simeq 43$\,day, respectively, in the observer's frame), which are longer
than the soft X-ray time scales detected in blazars and Seyfert galaxies. We find that the mixed OU process
approximates the light curves of the remaining 10 blazars better than the OU process. We derive limits on their long and
short characteristic time scales, and infer that their Fermi/LAT PSD resemble power-law functions.
We constrain the PSD slopes for all but one source in the sample. We find hints
for sub-hour Fermi/LAT variability in four flat spectrum radio quasars. We discuss the implications of our results
for theoretical models of blazar variability.
\end{abstract}

\keywords{black hole physics -- BL Lacertae objects: general -- galaxies: active -- galaxies: jets -- gamma rays: galaxies}

\section{Introduction}

A significant fraction of active galactic nuclei (AGN)
produce powerful relativistic jets, which are prominent
sources of non-thermal radiation. In blazars, where one
of the jets is closely aligned with our line of sight,
this non-thermal radiation component is relativistically
boosted to the point that it easily outshines the entire
host galaxy. Spectral energy distributions of many blazars
peak in the gamma-ray band, making data from the Fermi
Large Area Telescope (Fermi/LAT) essential for studying the jet
physics.

The gamma-ray emission of blazars is well known for its
strong and incessant variability, indicating the complex
structure of the underlying dissipation and particle
acceleration sites. It proved to be very difficult to provide
a satisfactory statistical description of these variations
that would be useful for constraining such basic parameters
as, e.g., the location and size of the dissipation sites along the jet.
Before the launch of Fermi, thorough blazar variability studies were
impaired mainly due to the lack of $\gamma$-ray blazar
monitoring data, and robust statistical methods to model the variable
blazar emission.

The Fermi/LAT instrument has been performing continuous observations
of the $\gamma$-ray sky since 2008, which provided good quality
light curves of a sample of bright blazars, and boosted the blazar
$\gamma$-ray variability studies. 
Recently, it has been demonstrated that the $\gamma$-ray power spectral
densities (PSD) of blazars appear to be in the form of a power-law function
(e.g. Abdo et al. 2010; Nakagawa \& Mori 2013, hereafter NM13),
indicating a stochastic nature of the high energy blazar variability.
On the other hand, the highest flux states
of blazars are commonly described as flares, defined using the concept of the
flux doubling and halving time scales (e.g. Nalewajko 2013, Saito et al. 2013).
Such an approach suggests that
flares have an origin that is distinct from that of the bulk of
the $\gamma$-ray blazar variability.

An important characteristic of a variability process is a frequency or a time scale
at which the properties of its PSD (e.g. the slope) change. This time
scale may be related to, e.g., the size of the emitting region, and used to
constrain the process triggering the variability. However, the featureless
power-law like Fermi/LAT blazar PSD have been so far preventing this kind of
inference, with 3C 454.3 being the only blazar with a
$\gamma$-ray PSD break reported in the literature. Ackermann et al. (2010) 
performed a PSD and structure function analyses of a 120\,day flaring section of the
Fermi/LAT light curve of this source. They revealed a break at frequency
corresponding to a specific time scale $t\sim6.5$\,day. Subsequently, NM13
found a break frequency
corresponding to $t\sim7.9$\,days in the first
4-year Fermi/LAT light curve of 3C 454.3. Ackermann et al. (2010) cautioned that
their PSD break may not indicate a characteristic time scale but a
frequency at which two PSD components become equally strong. NM13 interpreted
their characteristic time scale in terms
of the internal shock model for the $\gamma$-ray blazar emission, in which
blob ejecta collide in the internal shock of the blazar jet (Kataoka
et al. 2001). This assumption allowed them to estimate the black hole
mass in 3C 454.3 to be in the $10^8$--$10^{10}$\,M$_{\odot}$ range.
The PSD slopes below and above the breaks estimated by Ackermann et al. (2010)
and NM13 as well as the location of the PSD breaks were inconsistent with each other
which may suggest either a non-stationarity of the variability process
in this source, or indeed distinct variability properties of the $\gamma$-ray
flares.

The methods relying on the PSD extraction require that the light curves are
uniformly sampled, or a number of biases are introduced and have to be
accounted for, which is not trivial.  Models of variability applied
directly to the light curves avoid these biases. Kelly et al. (2009, 2011)
developed and advocated for stochastic models for luminosity fluctuations of accreting
black holes, motivated by PSD proportional to $1/f^{\alpha}$ with one or two breaks
that have been commonly observed in X-rays
in the black hole binaries (e.g. Pottschmidt et al. 2003, Axelsson et al. 2005, Belloni
et al. 2005, Reig et al. 2013),
and in optical and X-ray bands in AGN (e.g. Markowitz et al. 2003, Kelly et al. 2009).
The models of Kelly et al. are based on the Ornstein-Uhlenbeck process
or a linear superposition of the OU processes. Thus,
they explicitly assume a power-law PSD with one or two characteristic time scales.
Kelly et al. derive the
likelihood function for their statistical models and perform statistical
inference within a Bayesian framework. This allows them to obtain the
probability distributions of the model parameters, such as the characteristic
time scales and the slopes of the intermediate part of the PSD, given the
data. They fully account for the measuring errors, irregular sampling,
red noise leak, and aliasing. In addition, direct modeling of the light curves
allows to combine easily different sampling time scales.
All these advantages
make the Kelly et al. models particularly attractive for constraining the
PSD of AGN in all energy bands.

In this paper we apply the stochastic models of Kelly et al. to the first 4-year
Fermi/LAT light curves of 13 bright blazars. This is the first systematic analysis of
the $\gamma$-ray blazar light curves in the time domain using the
parametric methods which are not sensitive to the observational
biases. The paper is organized as follows. In Section~\ref{sec:sample} we describe
the blazar sample and data reduction procedure. The models are summarized in Section~\ref{sec:model}.
In Section~\ref{sec:results} we present our results on the derived constraints on the PSD
parameters for the sources in our sample. In Section~\ref{sec:discussion} we 
discuss our findings on the $\gamma$-ray blazar variability and perform a comparison with 
with the X-ray variability properties of blazars and non-blazar AGN. We formulate our
conclusions in Section~\ref{sec:conclusions}.

\section{Fermi/LAT sample}
\label{sec:sample}

We extract $\gamma$-ray light curves for a sample of 13 bright blazars using the
first 4 years of the publicly available Fermi/LAT data (MJD 54682 -- 56143).
The sample consists of 8 flat spectrum radio quasar type sources (FSRQ) and 5 BL Lacs
(Table~\ref{tab:sample}) that were among the
brightest sources in the 2FGL catalog (Nolan et al. 2012) and remained bright
during the 3rd and 4th years of the Fermi mission as indicated by the Monitored
Source List (e.g., we excluded AO 0235+164). The 100\,MeV--300\,GeV $\gamma$-ray
fluxes were calculated in standard unbinned maximum likelihood
analysis using the Science Tools software package v9r27p1, instrument response
function P7SOURCE\_V6, Galactic diffuse emission model gal\_2yearp7v6\_v0, isotropic
background model iso\_p7v6source, 10\,deg region of interest, and 2FGL-based 15\,deg
background source model.

\begin{figure*}
\begin{center}
\includegraphics[height=0.25\textheight]{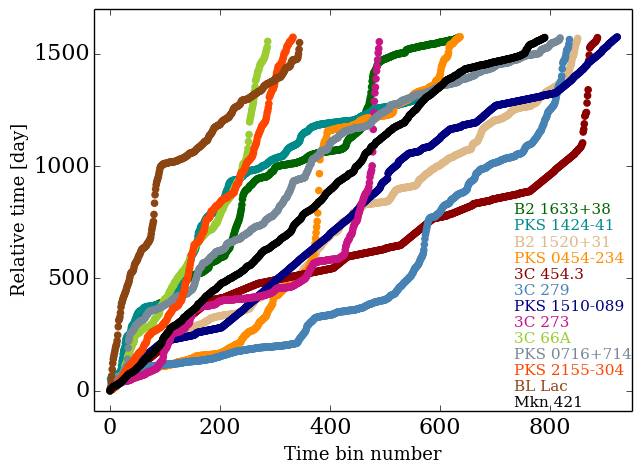}
\caption{Relative time as a function of the time bin number. The time bin sizes are in the $\Delta t \simeq 0.02$--50\,day range. The time resolution of several light curves is close to uniform. Other light curves contain periods with crude time resolution corresponding to the low flux states (vertical rises), which separate fine time resolution periods with high flux.}
\label{fig:dt}
\end{center} 
\end{figure*}

Time bins were selected in an adaptive algorithm by dividing each bin (starting
from 200\,day bins) into equal halves as long as the likelihood analysis for each half
satisfied the minimum test statistic criterium ${\rm TS} \ge 25$. The resulting time resolution
of the light curves varies in the $\Delta t \simeq 0.02$--50\,day range. In Figure~\ref{fig:dt}
we plot the relative time of each light curve as a function of the time bin index. In
this representation a uniformly binned light curve would be given by a linear
relation. It can be seen that the time bin distribution in some light curves (PKS 1510-089, B2 1520+31
and Mrk 421) reseambles closely a uniform distribution. On the other hand, 
light curves of PKS 0454-234, 3C 454, and 3C 273 contain time periods with crude time
resolution corresponding to the low flux states of the sources reflected by vertical
rises in Figure~\ref{fig:dt}, which separate periods of fine sub-day time resolution
associated with the high flux states.

\section{Modeling}
\label{sec:model}

We model the Fermi/LAT light curves of the sources in our sample as
a single Ornstein-Uhlenbeck (OU) process (Kelly et al. 2009), or a linear superposition of
the OU processes (hereafter sup-OU, Kelly et al. 2011). In this section we summarize briefly the
main properties of these processes, and we refer the reader to Kelly et al. (2009, 2011)
for the full details of the models construction and fitting procedures.

The OU process, $X(t)$, is defined
by the following stochastic differential equation:
\begin{equation}
dX(t) = -\omega_0 ( X(t) - \mu) dt + \varsigma dW(t), \omega_0, \varsigma > 0. 
\end{equation}
The parameters of the process are the characteristic frequency, $\omega_0$; the mean value of the process, $\mu$; and
the amplitude of the driving noise process, $\varsigma$, being the rate at which variability power is injected into the
stochastic process $X(t)$. The term $W(t)$ denotes a Wiener process (i.e., a Brownian motion), and its derivative is
white noise. The power spectrum of an OU process is flat for frequencies $\omega \ll \omega_0$ and decays as
$1/\omega^2$ for frequencies $\omega \gg \omega_0$. Hence, $\omega_0$ may be associated with a characteristic time scale
$\tau_{\rm 0}=1/\omega_0$ related with a change in the properties in the PSD of the OU process. The OU process
proved to provide an adequate description of the optical light curves of AGN (Kelly et al. 2009; Koz{\l}owski et al. 2010;
MacLeod et al. 2010; Zu et al. 2013; Andrae et al. 2013), at least on time scales longer than a $\sim$3 months
(Mushotzky et al. 2011).

A process defined as a combination of the OU processes,
\begin{equation}
 Y_M(t) = \mu + \sum_{i=1}^{M} c_i X_i(t),
\end{equation}
where $c_1$,..., $c_M$ is a set of mixing weights, and $X_1(t)$,..., $X_M(t)$ is a set of the OU processes, allows to model
light curves with more complex PSD, e.g. those that are flat below a low-frequency break, $\omega_1$, decay as
$1/\omega^\alpha$ with $\alpha=0$--2 above the low-frequency break, and then steepen to $1/\omega^2$ above a
high-frequency break, $\omega_2$, as observed e.g. in the X-ray PSD of nearby
AGN and black hole X-ray binaries (e.g. Kelly et al. 2011, Axelsson et al. 2005). We define the short and long term characteristic time
scales related to the two characteristic frequencies as $\tau_S = 1/\omega_2$ and $\tau_L = 1/\omega_1$.
The characteristic time scales that we sample for are between $\tau_{\rm max}$ and $\tau_{\rm min}$
chosen to be 100 (10) times the length of the time series, and one-hundredth (one-tenth) the smallest time spacing
in the time series in the OU (sup-OU) model. Throughout the paper it holds that a Fourier frequency
$f = t^{-1} = \omega/2\pi = (2\pi \tau)^{-1}$, where $t$ is a specific time scale. Thus, our study covers a wide range
of characteristic time scales corresponding to the $f \simeq 10^{-11}$--10$^{-3}$\,Hz and $f \simeq 10^{-10}$--10$^{-4}$\,Hz
range in Fourier frequency in the OU and sup-OU models, respectively.

Uttley et al. (2005) argued that the X-ray light curves of X-ray binaries and AGN, $x(t)$, are formally non-linear, and
can be described by the simple transformation $x(t) = \exp[l(t)]$, where $l(t)$ is a linear, Gaussian time series. The
observational evidence supporting this conclusion include the linear rms-flux relation observed on all time scales in
the X-ray variations of GBHs and AGNs (Uttley \& McHardy 2001), and the log-normal distribution of fluxes for Cygnus
X-1 (Uttley et al. 2005). Thus, we model the natural logarithm of the Fermi/LAT fluxes in order to investigate whether the
$\gamma$-ray light curves of AGN share the properties of the optical and X-ray AGN light curves.

The error bars derived in this work are representative of the 90\% confidence regions, while the upper and lower limits
represent the 99\% confidence regions. While reporting the results, we use logarithms to base 10 and calculate time scales in the observer's frame, unless otherwise stated.

\section{Results}
\label{sec:results}

\subsection{The OU process}

\begin{figure*}
\begin{center}
\includegraphics[height=0.3\textheight]{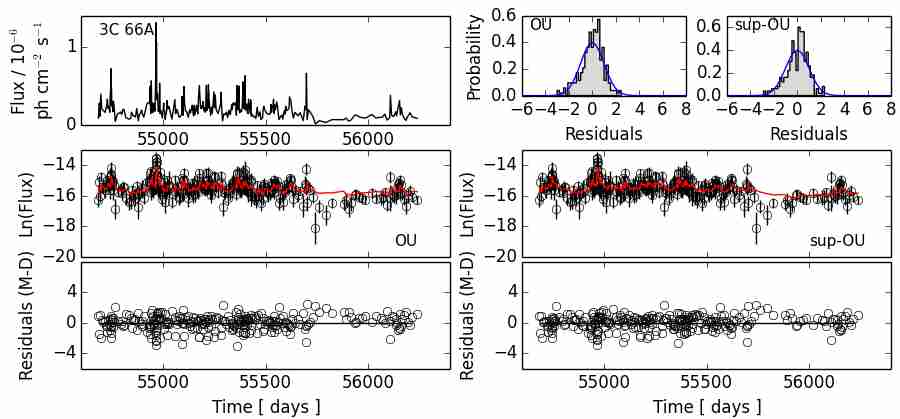}
\includegraphics[height=0.3\textheight]{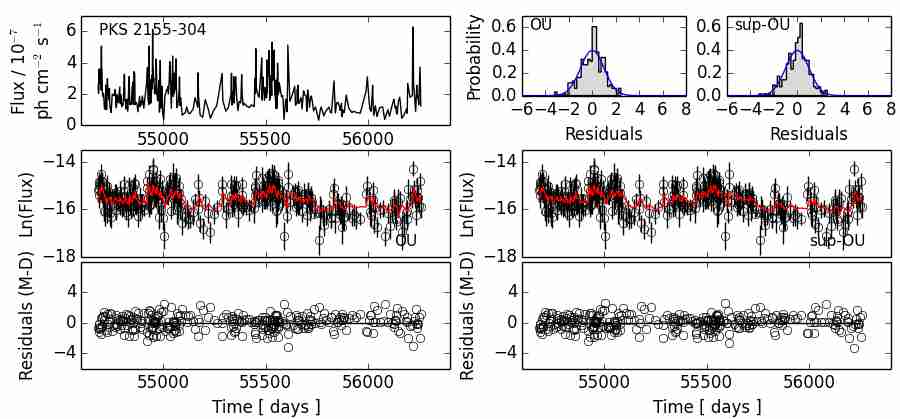}
\includegraphics[height=0.3\textheight]{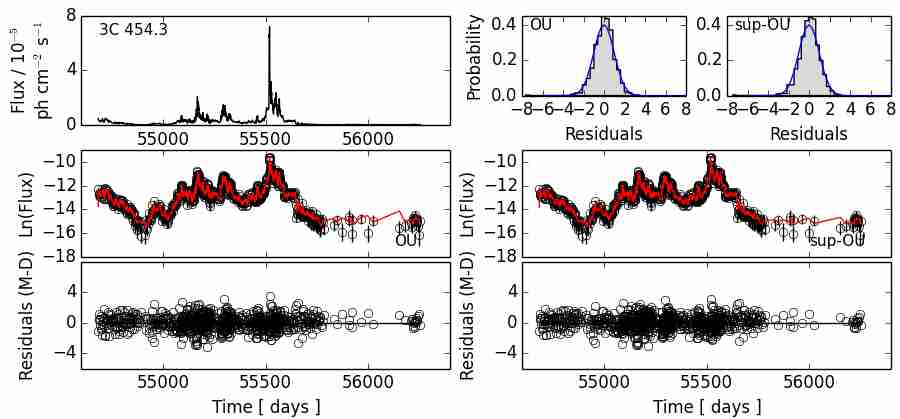}
\caption{Results of modeling for 3C 66A, PKS 2155-304, and 3C 454.3. The sub-panels show for each source the Fermi/LAT light curves (top left); natural logarithm of the light curves, the OU/sup-OU model corresponding to the highest posterior likelihood, and the model standardized residuals (bottom left/bottom right); the histograms of the standardized residuals compared to the expected standard normal distribution (top right). Neither model can be favored based on DIC ($\Delta{\rm DIC}<10$ in each source).}
\label{fig:fits-ou}
\end{center}
\end{figure*}

\begin{figure*}
\begin{center}
\includegraphics[height=0.3\textheight]{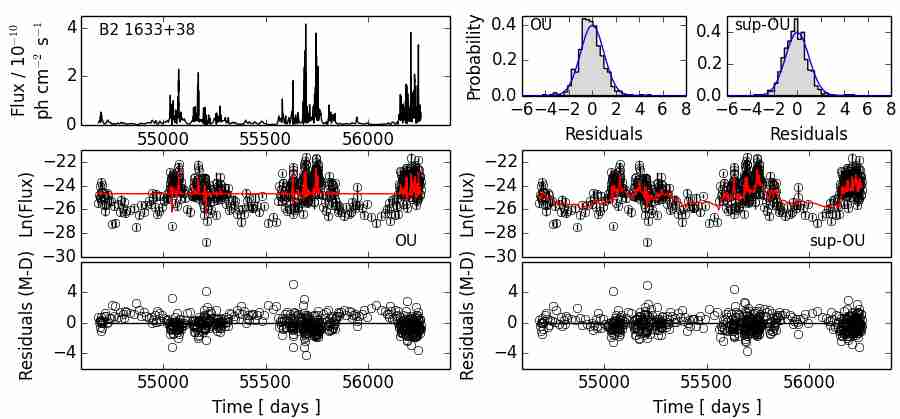}
\includegraphics[height=0.3\textheight]{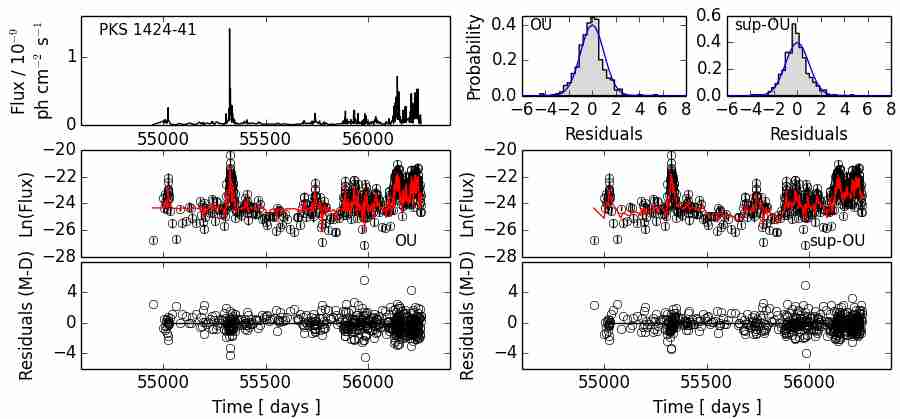}
\includegraphics[height=0.3\textheight]{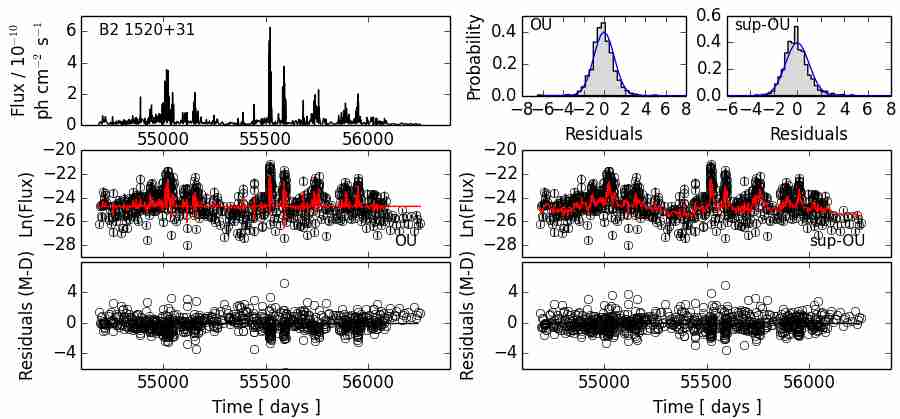}
\caption{Same as Figure~\ref{fig:fits-ou} for B2 1633+38, PKS 1424-41 and B2 1520+31.
The sup-OU model is favored based on DIC ($\Delta{\rm DIC}>10$).}
\label{fig:fits-supou1}
\end{center}
\end{figure*}

\begin{figure*}
\begin{center}
\includegraphics[height=0.3\textheight]{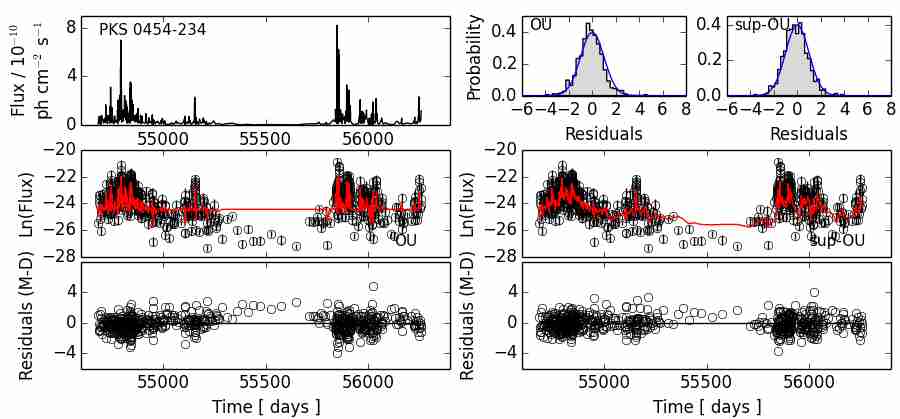}
\includegraphics[height=0.3\textheight]{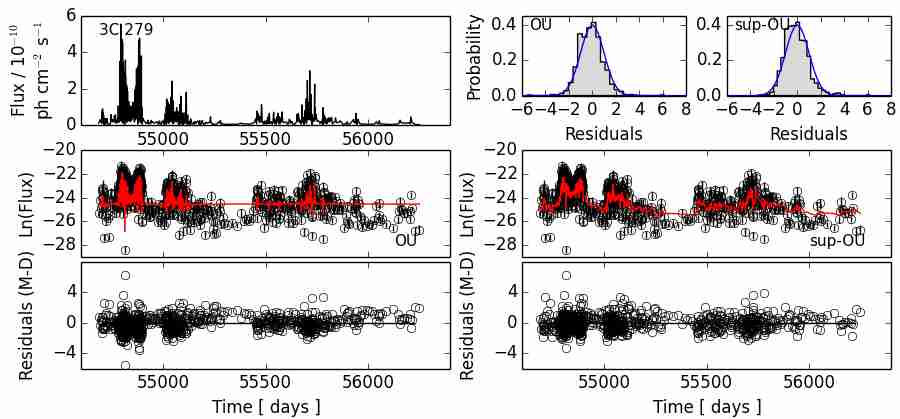}
\includegraphics[height=0.3\textheight]{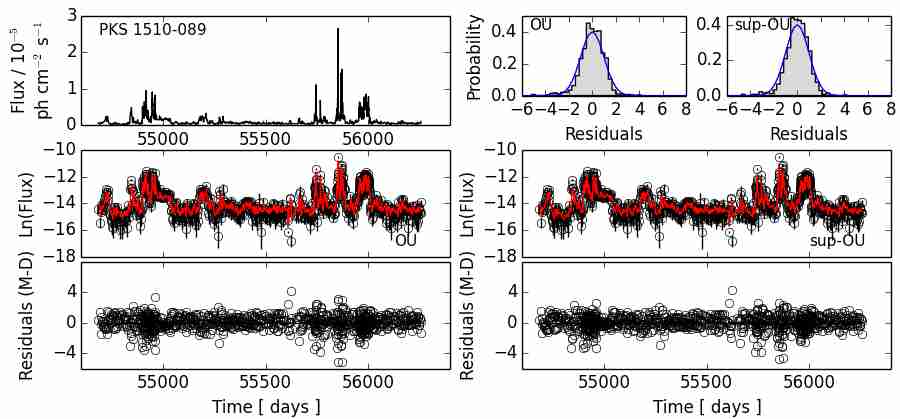}
\caption{Same as Figure~\ref{fig:fits-ou} for PKS 0454-234, 3C 279 and PKS 1510-089.
The sup-OU model is favored based on DIC ($\Delta{\rm DIC}>10$).}
\label{fig:fits-supou2}
\end{center}
\end{figure*}

\begin{figure*}
\begin{center}
\includegraphics[height=0.3\textheight]{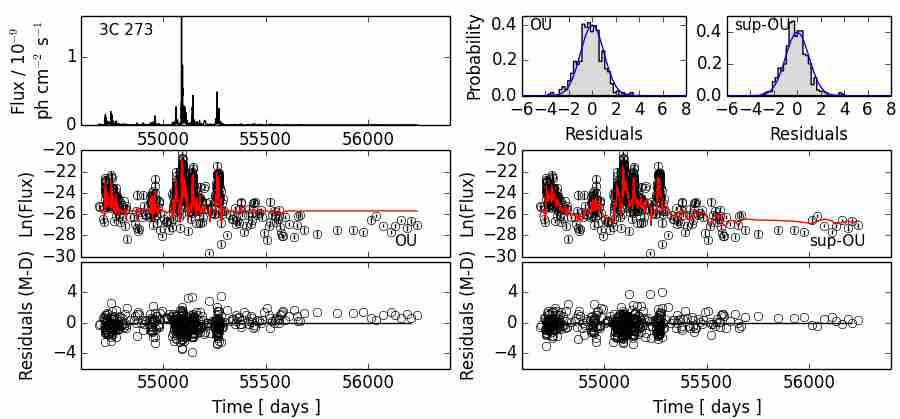}
\includegraphics[height=0.3\textheight]{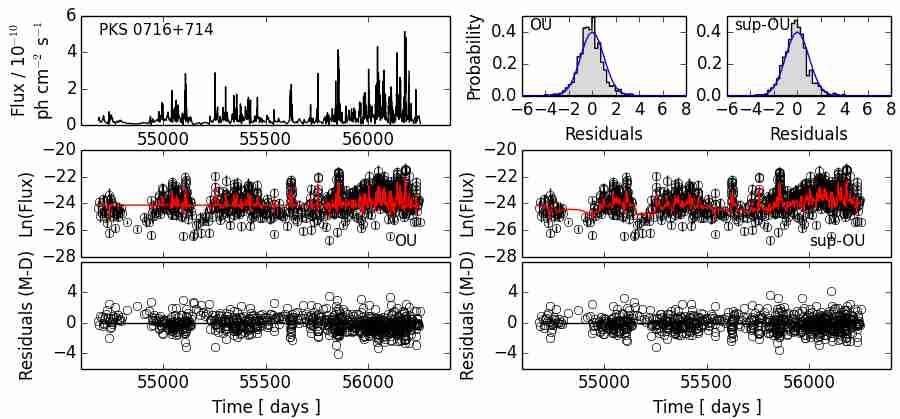}
\includegraphics[height=0.3\textheight]{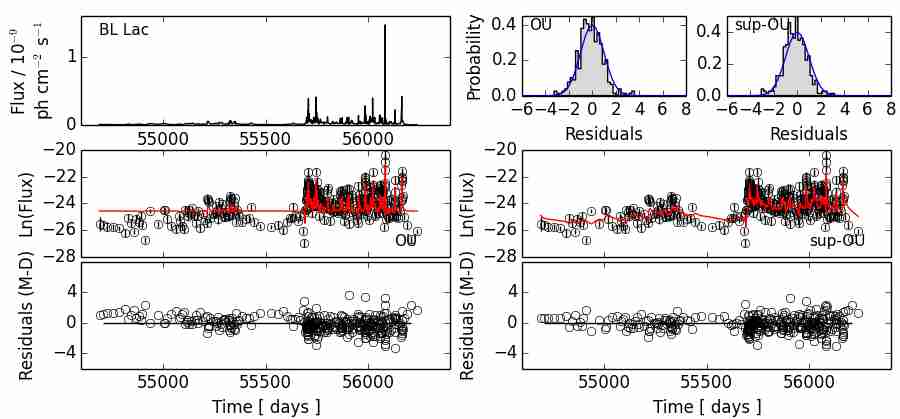}
\caption{Same as Figure~\ref{fig:fits-ou} for 3C 273, PKS 0716+714 and BL Lac.
The sup-OU model is favored based on DIC ($\Delta{\rm DIC}>10$).}
\label{fig:fits-supou3}
\end{center}
\end{figure*}

\begin{figure*}
\begin{center}
\includegraphics[height=0.3\textheight]{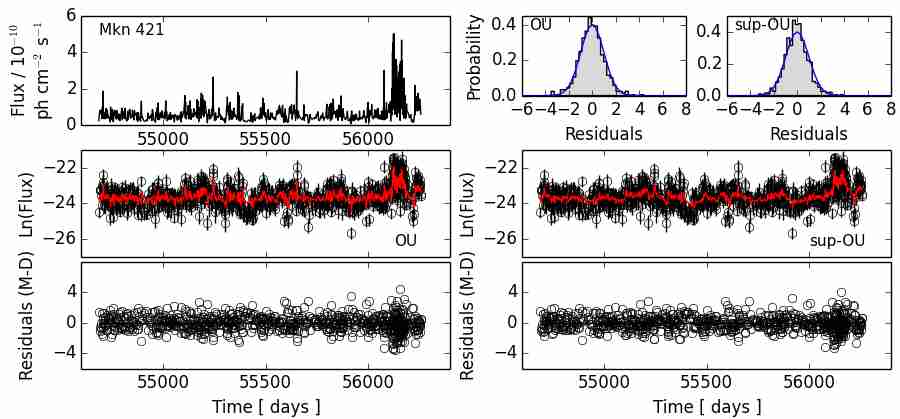}
\caption{Same as Figure~\ref{fig:fits-ou} for Mkn 421.
The sup-OU model is favored based on DIC ($\Delta{\rm DIC}>10$).}
\label{fig:fits-supou4}
\end{center}
\end{figure*}

\begin{figure*}
\begin{center}
\includegraphics[width=0.9\textwidth]{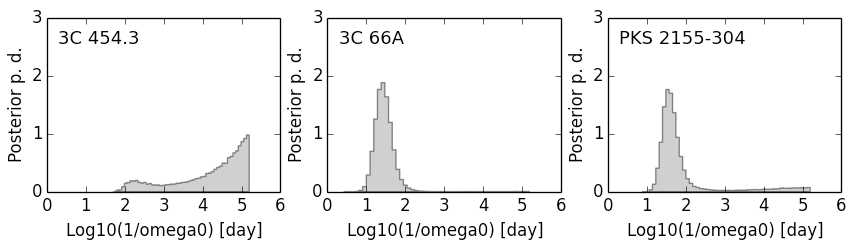}
\caption{Posterior probability distributions of the logarithm of the characteristic time scale, $\tau_0 = 1/\omega_0$, resulting from application of the OU process to the Fermi/LAT light curves of 3C 454.3, 3C 66A, and PKS 2155-304.}
\label{fig:omega0}
\end{center}
\end{figure*}

First, we modeled the Fermi/LAT light curves of the blazars in our sample with the OU process. We plot the light curves,
the model realizations corresponding to the highest posterior likelihood, the standardized model residuals and their histograms compared to the expected standard normal distribution in
Figures~\ref{fig:fits-ou}--\ref{fig:fits-supou4} (left). 
The sample posterior distribution functions of the characteristic time
scale, $\tau_{\rm 0}$ are presented in Figure~\ref{fig:omega0}. In Table~\ref{tab:sample} we list the
median characteristic
time scales and their uncertainties for all sources. The OU modeling gives the characteristic time scales in the
$\tau_{\rm 0} \simeq 0.7$--43\,day range (Table~\ref{tab:sample}). In the case of 3C 454.3 the characteristic time
scale is not constrained, and we derive the lower limit on its value,
$\log [\tau_{\rm 0} / {\rm day}] > 2.0$.

\begin{figure*}
\begin{center}
\includegraphics[height=0.9\textheight]{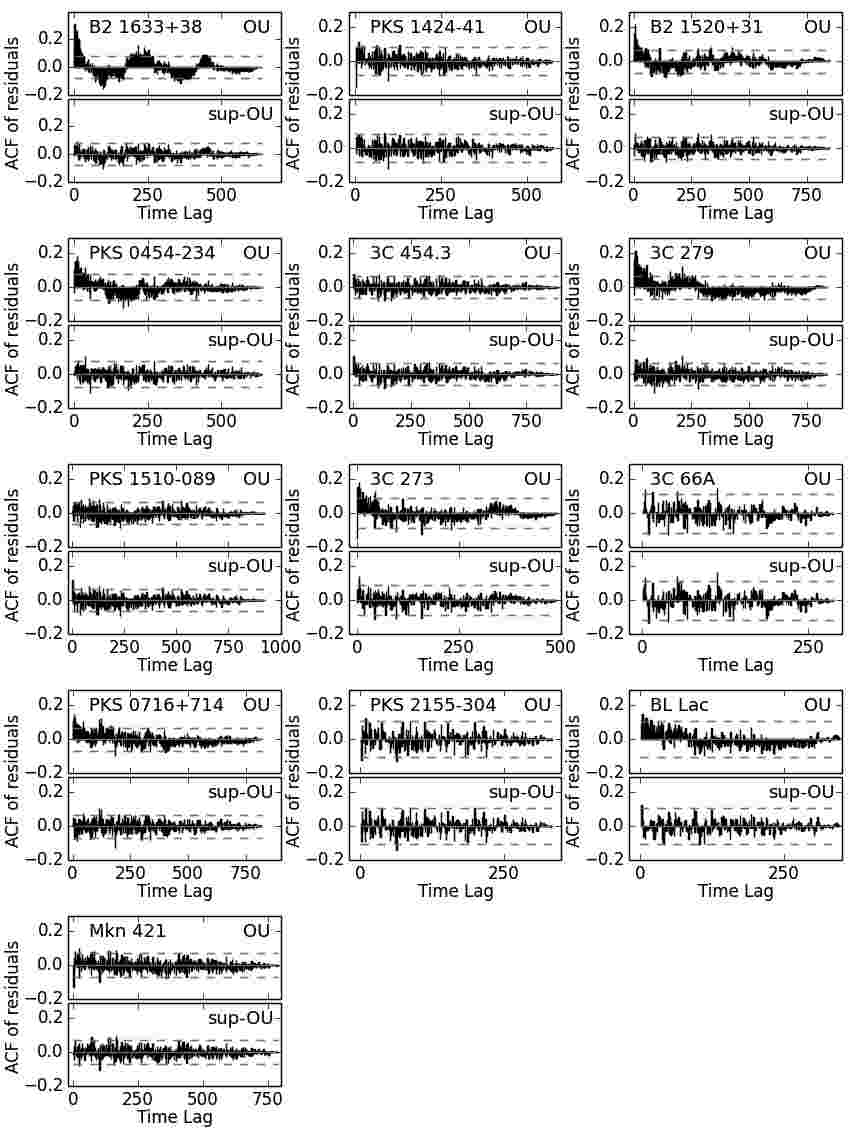}
\caption{Auto-correlation function of the standardized residuals in the OU and sup-OU models. The value at lag 0 is not plotted. The dashed lines represent the 95\% confidence limit on the auto-correlation function assuming the white noise process.}
\label{fig:acf}
\end{center}
\end{figure*}

It is apparent that in general the OU process does not fit well the studied light curves. In particular,
the model experiences systematic difficulties to fit the data during the quiet time periods when the assumed condition
that ${\rm TS} \ge 25$ forces the adaptive time bins
$\Delta t$ to be larger than a few days (see e.g. Figures~\ref{fig:fits-supou1}--\ref{fig:fits-supou3}, left).
To evaluate the quality of the fits, we compare the distributions of the standardized fit residuals with the standard
normal distribution. In addition for each source we calculate the auto-correlation function of the standardized
residuals (Figure~\ref{fig:acf}). It can be observed that while the distribution of the residuals approximate the standard normal distribution reasonably well, the auto-correlation functions deviate in several sources (e.g. B2 1633+38) from that expected for the white noise process. Thus, we proceed to testing the sup-OU model on the Fermi/LAT light curves of blazars.

\subsection{The sup-OU process}

In the next step we considered the superposition of the OU processes (sup-OU) and applied it to the Fermi/LAT blazar
light curves. We observed a significant improvement to the quality of the fits in the low-flux periods, compared to
the case of the OU modeling. 
The distributions of the standardized residuals assuming the sup-OU process can be compared with the OU model fit residuals
and the standard normal distribution (Figures~\ref{fig:fits-supou1}--\ref{fig:fits-supou4}, right). 
The auto-correlation functions of the standardized residuals assuming
the sup-OU process indicate that, indeed, in majority of sources the sup-OU residuals approximate the white noise process better than the OU process residuals (Figure~\ref{fig:acf}).
 
To evaluate formally what model is more successful in describing the Fermi/LAT blazar light curves, we computed the deviance
information criterion (DIC, Spiegelhalter et al. 2002). DIC is used in Bayesian model selection problems where the posterior distributions of
the models have been obtained by Markov chain Monte Carlo simulations. Models with smaller DIC are preferred to models
with larger DIC. It is accepted that $\Delta({\rm DIC})\gtrsim10$ is a difference substantial enough to prefer the
model with smaller DIC. We use the following definition of the DIC.
\begin{equation}
\label{eq:dic}
 {\rm DIC} = p_V + \bar{D},
\end{equation}
where $p_V = 0.5 \times {\rm Var}[D(\theta)]$ describes the effective number of model parameters and hence the model
complexity; while $D(\theta)$ is the deviance defined as $D(\theta) = -2\log[p(y|\theta)]$, where $p(y|\theta)$
is the model likelihood function, $y$ are the data, and $\theta$ denotes the parameters in the model.
The second term in Equation~\ref{eq:dic}, $\bar{D}$, is the expected value of the deviance.

The comparison of DIC computed for the two models (Table~\ref{tab:sample}) indicates that the sup-OU model is preferred
over the OU model in all sources except 3C 454.3, 3C 66A, and PKS 2155-304 in which
$\Delta{\rm DIC} = {\rm DIC}_{\rm supOU} - {\rm DIC}_{\rm OU} < 10$. Thus, for these three sources we 
accept the results of the OU modeling, while for the remaining sources we assume the results of the sup-OU modeling.

\begin{figure*}
\begin{center}
\includegraphics[height=0.9\textheight]{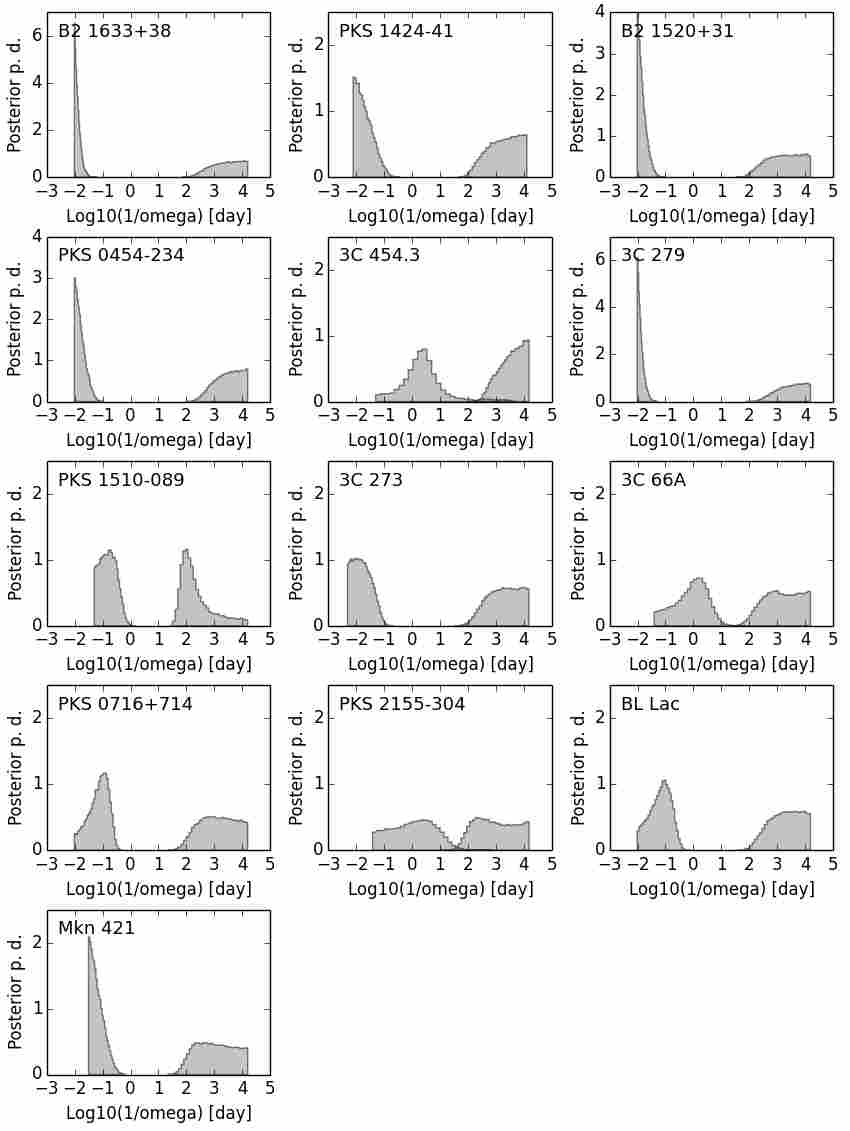}
\caption{Collection of the posterior probability distributions of the logarithm of the short and long characteristic time scales resulting
from application of the sup-OU process to the Fermi/LAT light curves.}
\label{fig:o1o2}
\end{center}
\end{figure*}

For each source, we plot the posterior probability distributions of the short and long characteristic time scales
resulting from the sup-OU model in Figure~\ref{fig:o1o2}. 
We notice that in many cases the posterior probability distributions of the characteristic time scales deviate
from a Gaussian distribution. It can be inferred that while formally we can derive the median long and short
characteristic time scales and their 90\% confidence intervals in all sources, they may depend on our choice
of $\tau_{\rm min}$ and $\tau_{\rm max}$.The least deviated shapes are those representing the long characteristic
time scale in PKS 1510-089; and the short characteristic time scales in 3C 454.3, and (to a lesser degree) in 3C 66A, PKS 0716+714, and BL Lac. 
Thus, we provide the median $\tau_{\rm L}$ and/or $\tau_{\rm S}$ and their 
uncertainties for these five cases, and the relevant limits for all sources 
(Table~\ref{tab:sample}). The upper limits on $\tau_{\rm S}$ can be compared with the smallest time spacing in the respective
light curves, $\Delta t$ (see Table~\ref{tab:sample}), and the shortest sampled time scale, $\tau_{\rm min} = \Delta t / 10$.

We present the posterior probability distributions of the slopes, $\alpha$, characterizing the intermediate part of the
PSD between $\tau_{\rm L}$ and $\tau_{\rm S}$ in Figure~\ref{fig:slope}. The distributions peak close to $\alpha \simeq 1$ for
the majority of the sources. The individual values and their uncertainties are listed in
Table~\ref{tab:sample}. The slopes do not depend on the blazar class (BL Lac vs. FSRQ) or redshift (Figure~\ref{fig:results}, right;
left). With the sup-OU model, we constrain the PSD slopes in all sources, except 3C 454.3 for which we find $\alpha < 0.43$.

\section{Discussion}
\label{sec:discussion}

We analyzed the variability properties of 13 bright blazars with the most complete first 4-year Fermi/LAT light
curves available from the satellite archive. We used the stochastic models for luminosity fluctuations developed
in Kelly et al. (2009, 2011). The models
rely on computing the likelihood function and performing Bayesian inference to access the probability distributions
of the variability parameters such as, e.g., the characteristic time scales, given the data. The models implement the OU
(Ornstein-Uhlenbeck) process leading to a bending PSD, and the superposition of the OU processes leading to a doubly
bent PSD. These shapes of PSD are typically observed in radio-quiet AGN in the optical and X-ray bands, and in X-rays
in the black hole binaries across different spectral states. Our approach provides an alternative to the methods
relying on the PSD extraction known to suffer from a number of observational biases due to irregular sampling, read noise
leak, and aliasing. Moreover, our approach allows for non-uniform binning of the light curves. Hence, in this
work we binned the light curves adaptively in order to avoid compromising the time resolution in high flux periods
for the sake of uniform binning over the entire observing period. In addition, in our modeling we sampled a wide range of
temporal frequencies, up to 8 orders of magnitude starting at $f_{\rm min} \simeq 10^{-11}$--$10^{-10}$\,Hz. Thus, we
improved significantly over the earlier Fermi/LAT blazar variability studies (e.g., $f \simeq 0.003$--0.2\,Hz,
Abdo et al. 2010; $f \simeq 10^{-7}$--$10^{-5}$\,Hz, NM13).

\subsection{Characteristic time scales in blazars}

The main result of our study is the consistency of the 4-year Fermi/LAT blazar activity with stochastic processes.
The mixed OU process was prefered in 10 blazars from our sample, while in the 3 remaining blazars (two BL Lacs, 3C 66A
and PKS 2155-304; and one FSRQ 3C 454.3) the single OU process and the mixed OU process resulted in fits
of equally good quality.

The posterior probability distributions of the characteristic time scales from the OU process
allowed us to constrain the Fermi/LAT characteristic time scales in two BL Lac sources
($\log [\tau_{\rm 0} / {\rm day}] = 1.39^{+0.43}_{-0.29}$ in 3C 66A, and
$\log [\tau_{\rm 0} / {\rm day}] = 1.63^{+2.79}_{-0.35}$ in PKS 2155-304). This is the
first time that the characteristic time scales have been observed in the Fermi/LAT light curves of blazars other
than 3C 454.3. In 3C 454.3 the OU modeling resulted only in a lower limit, $\log [\tau_{\rm 0} / {\rm day}] > 2.0$,
which was inconsistent with the NM13 result. NM13 found that the PSD of 3C 454.3 can be characterized with $\alpha\simeq1$
and $\alpha\simeq3$ below and above the PSD break, respectively. The high frequency slope significantly steeper than 
$\sim$2 assumed in our modeling might have led to poor constraints on the characteristic time scales
in this source. Characteristic time scales found through the OU modeling in 3C 66A and PKS 2155-304 fall outside of
the frequency range studied by NM13. We conclude that perhaps the method of the PSD construction used by NM13 was not
sensitive enough to detect these BL Lac characteristic time scales in the Fermi/LAT light curves.

With regard to the sup-OU modeling, we reported the relevant limits on the characteristic time scales. In five cases
with the posterior probability distributions resembling most closely a Gaussian distribution ($\tau_{\rm S}$ in 3C 454.3,
3C 66A, PKS 0716+714, and BL Lac; $\tau_{\rm L}$ in PKS 1510-089) we provided additionally the median values and their
uncertainties. However, we cautioned that they might be affected by our choice of
the lowest and highest sampled time scales, $\tau_{\rm min}$ and $\tau_{\max}$.
The upper limits on the short characteristic time scales obtained in our studies hint toward a sub-hour $\gamma$-ray time
variability in at least 4 blazars: B2 1633+38, B2 1520+31, PKS 0454-234, and 3C 273, all belonging to the FSRQ class
(Figure~\ref{fig:results}, left).
Sub-hour variability time scales have been reported previously in another FSRQ, PKS 1510-089, in the context of the $\gamma$-ray flux
doubling/halving time scales, rather than the characteristic time scales, in the study of the $\gamma$-ray flares
in this source (Saito et al. 2013). Our adaptive light curves do not sample the sub-hour time scales due to
a conservative TS constraint in each time bin. However, even with less constraining condition on TS it would be difficult
to obtain sub-hour time resolution for a prolonged time required by our study, given the variable nature of blazars,
as well as the characteristics of the Fermi/LAT instrument. Obtaining sub-hour time resolution in the Fermi/LAT light
curves is feasible only for the highest $\gamma$-ray flux periods observed in the $\gamma$-rays. Yet, the results might
be affected by the poorly understood systematic errors related to the Fermi/LAT data analysis on sub-orbital time scales.
The lower limits on the long characteristic time scales indicate that observations conducted for a period of time longer
than 4 years considered here are needed in order to constrain $\tau_{\rm L}$.

Ackermann et al. (2010) computed a PSD and structure function of a 120\,day long section of Fermi/LAT light
curve of 3C 454.3, during which a powerful flare was recorded (MJD 55,120--55,260). They reported a 6.5\,day time scale
for this time period. We tested our models on the light curve of 3C 454.3 spanning the same range in MJD.
Both the OU and sup-OU models resulted in fits of comparable quality ($\Delta{\rm DIC} \simeq 3$).
However, the $\tau_{\rm S}$ and $\tau_{\rm L}$ distributions obtained through the sup-OU modeling were poorly
separated indicating at most a presence of one characteristic time scale in the light curve. The OU modeling was
not successful in constraining the time scale and we obtained only a lower limit, $\log [\tau_{\rm 0} / {\rm day}] > 1.0$.
The OU limit on $\tau_{\rm 0}$ derived for the 120\,day light curve of 3C 454.3 and the 6.5\,day time scale
reported in Ackermann et al. (2010) are in disagreement, most probably due to the difference between
the PSD assumed in the OU model and that found in Ackermann et al. (2010; $\alpha\simeq1.40$ and $\alpha\simeq1.56$ below
and above the PSD break, respectively).

\subsection{Blazar PSD slopes}

\begin{figure*}
\begin{center}
\includegraphics[height=0.9\textheight]{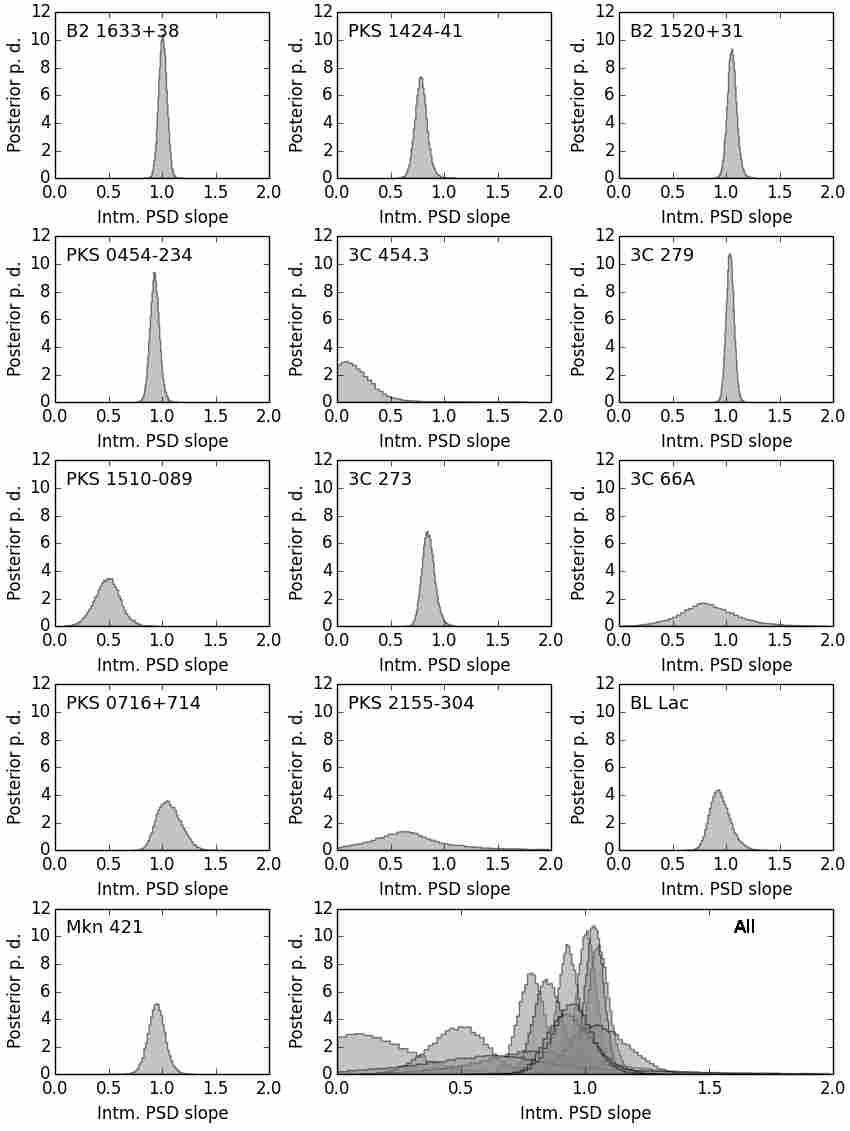}
\caption{Collection of the posterior probability distributions of the intermediate PSD slope, $\alpha$, resulting from
the application of the sup-OU process to the Fermi/LAT light curves. The bottom-right panel compares distributions of all sources.}
\label{fig:slope}
\end{center}
\end{figure*}

In 10 of 13 Fermi/LAT blazar light curves in our sample the sup-OU model was preferred over the OU model based on DIC.
Thus, the second important result of our study is that for the range of time scales sampled with the sup-OU model
(Figure~\ref{fig:results}, left), the underlying PSD
of these 10 sources resemble a power-law function proportional to $1/f^{\alpha}$ with $0<\alpha<2$, with majority of the slopes
clustering around unity (Figure~\ref{fig:slope}).
We were able to constrain the slopes $\alpha$ in all but one source, 3C 454.3, for which we found $\alpha < 0.43$.

Abdo et al. (2010) reported on an average Fermi/LAT PSD slope for their brightest FSRQs, $\alpha_{\rm FSRQ} = 1.5\pm0.2$,
and BL Lac's, $\alpha_{\rm FSRQ} = 1.7\pm0.3$. In general, the slopes derived for the individual sources in our sample
through the sup-OU modeling are shallower than those respective averages. In the FSRQ class, slopes of all our sources are
inconsistent with the Abdo et al. average. In the BL Lac class, only PKS 2155-304 and (marginally) BL Lac and Mkn 421 have
the slopes consistent with the Abdo et al. average at the 2$\sigma$ confidence level, but these are the three source with
the broadest posterior slope distributions in our sample (see Figure~\ref{fig:slope}).

We also compared our derived PSD slopes to those found by NM13 (Table~\ref{tab:sample}).
We found that in PKS 0454-234, 3C 279, 3C 66A, and PKS 2155-304 the two methods are consistent with each other
in terms of the derived PSD slopes within the provided error bars. In 3C 454.3, PKS 1510-089, 3C 273, BL Lac and Mkn 421
our slopes were shallower than those derived by NM13. In 3C 454.3 we suspect that the reason for the discrepancy is in
our assumption that the high frequency slope equals 2. If this slope is as steep as 3 reported in NM13 then our modeling
might result in a slope below the high frequency break that is flatter than the actual value in order to maintain
a satisfactory fit.


\subsection{Models of blazar variability}

The origin of blazar $\gamma$-ray variability seen in the Fermi/LAT light curves remains under discussion.
In our studies, we followed a stochastic approach to describe this variability motivated by its success 
in explaining the optical and/or X-ray variability of non-blazar accreting black holes. We demonstrated that
the OU and/or mixed OU processes approximate well the blazar $\gamma$-ray light curves. Our modeling implies
non-linearity of these blazar light curves, similarly to the case of the black hole binary and AGN X-ray light curves,
as argued by Uttley et al. (2005).

The interpretation of our models with regard to the optical
and/or X-ray black hole variability is along the line of the propagation class of models (e.g. Lyubarskii 1997, Kotov et al. 2001)
where perturbations originating at outer radii of an accretion disk propagate inwards and couple with each other producing
the observed broad PSD shapes. In particular, Kelly et al. (2011) showed that the mixed OU process is the solution to the linear
diffusion equation perturbed by a spatially correlated noise field. Within this framework the low-frequency break in
the PSD corresponds to the diffusion time scale in the outer region of the accretion flow, and the shape of the PSD above
the low-frequency break depends on the viscosity of the flow. An additional high-frequency break in the PSD may exist
if the noise field is spatially correlated. The high-frequency break corresponds then to the
time it takes a perturbation traveling at the viscous speed to cross the characteristic spatial scale of the noise field. 

\begin{figure*}
\begin{center}
\includegraphics[height=0.27\textheight]{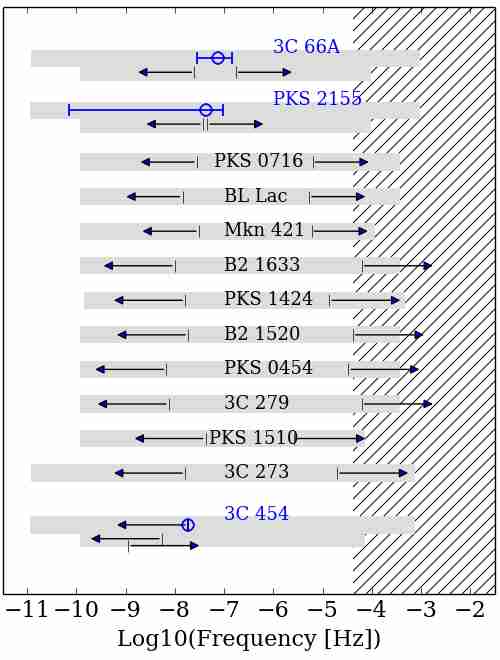}
\includegraphics[height=0.27\textheight]{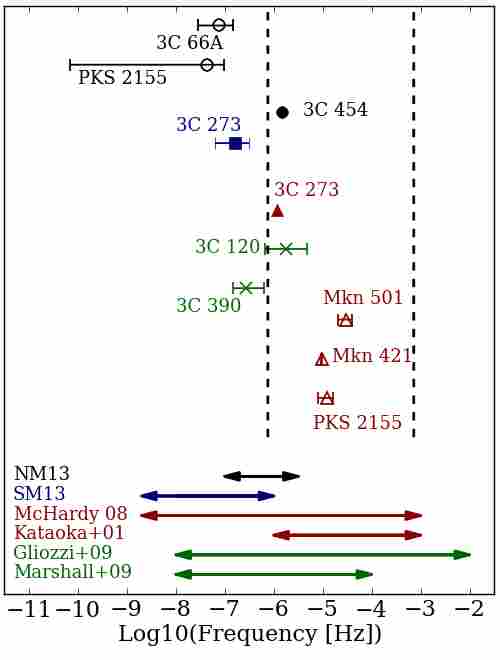}
\includegraphics[height=0.27\textheight]{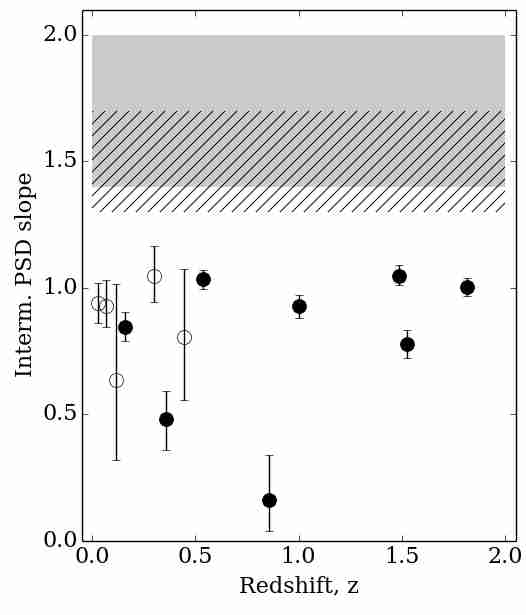}
\caption{Left: Constraints on the characteristic time scales in the Fermi/LAT blazar light curves modeled with our stochastic approach
(99\% confidence regions in the case of the limits, 90\% confidence regions otherwise).
Gray horizontal rectangles indicate the range of sampled frequencies. The hatched area marks the frequencies corresponding
to sub-hour time scales, $\tau < 1$\,hr.
In the case of 3C 66A, PKS 2155-304, and 3C 454.3 results of both the OU and sup-OU modeling are shown; the open circles indicate the OU model constraints.
Middle: Compilation of the blazar characteristic time scales across different energy bands from the literature.
Open symbols -- BL Lac sources, filled symbols -- FSRQ sources;
triangles -- soft X-rays, squares -- hard X-rays, circles -- $\gamma$-rays,
open circles -- this work; vertical dashed lines -- range of
$\tau_{\rm S}$ detected in the radio-quiet Seyfert galaxies in soft X-rays (Kelly et al. 2011); crosses -- detections of soft X-ray $\tau_{\rm S}$ in radio-loud AGN.
Arrows show the range of frequencies sampled by Kataoka et al. (2001; PKS 2155-304, Mkn 421, Mkn 501), Gliozzi et al. (2009; 3C 390.3),
Marshall et al. (2009; 3C 120), McHardy (2008; 3C 273, no error bars provided), Shimitzu \& Mushotzky (2013; 3C 273),
NM13 (3C 454.3, no error bars provided).
Right: Intermediate PSD slope as a function of redshift (filled circles - FSRQs, open circles - BL Lacs). The error bars represent
the 68\% (1$\sigma$) confidence intervals. The hatched and solid rectangles mark the 1$\sigma$ confidence limits on the
average PSD slopes in BL Lac and FSRQ sources, respectively (Abdo et al. 2010).}
\label{fig:results}
\end{center}
\end{figure*}

The short characteristic time scale corresponding to the high-frequency break in the black hole binary and AGN
X-ray PSD is an important variability parameter due to its correlation with
the black hole mass and mass accretion rate that holds over many
orders of magnitude (McHardy et al. 2006, Kelly et al. 2011).
Kelly et al. (2011) demonstrated that $\tau_{\rm S}$
obtained through direct modeling of the 2--10\,keV X-ray light curves of a sample of 10 nearby Seyfert galaxies (Sobolewska \& Papadakis 2009)
with the sup-OU process are in agreement with those derived using the PSD construction methods (Markowitz et al. 2003,
McHardy et al. 2007, Gonz{\'a}lez-Mart{\'{\i}}n \& Vaughan 2012).
However, blazar $\gamma$-ray emission is dominated by the processes related to the jet physics and beamed
due to relativistic effects, unlike the X-ray coronal emission of the radio-quiet AGN and black hole binaries. 
Nevertheless, it could be envisaged that the accretion disk variability structure is carried outwards in the jet, leading to the jet
variability with properties inherited from the innermost radii of the accretion disk/X-ray corona system (e.g. Lohfink et al. 2013). It would be then expected
that the X-ray and Fermi/LAT short characteristic time scales of blazars were consistent with those found in X-rays in radio-quiet and radio-loud AGN.
Thus, it is interesting to compare the blazar and non-blazar variability properties across different energy bands in order to
search for evidence in favor or against a common variability mechanism (Figure~\ref{fig:results}, middle; note differing
frequency bands sampled by the respective studies).

Blazar PSD have been studied in only several sources in energy bands other than the $\gamma$-rays. 
Soft X-ray blazar PSD (here below 20\,keV) were computed for three BL Lac
sources (Mkn 421 and PKS 2155-304 also included in our Fermi/LAT sample, and Mkn 501) by Kataoka et al. (2001), and for FSRQ sources
by McHardy (2008; 3C 273), Chatterjee et al. (2008; 3C 279).
Characteristic time scales corresponding to the high-frequency PSD break were detected in all of these
sources with the exception of 3C 279.
The sampled frequency bands and detected PSD breaks are indicated in Figure~\ref{fig:results} (middle).
In general, these studies demonstrated that in soft X-rays blazars show high-frequency breaks and
PSD shapes below these breaks that are largely consistent with those of Seyfert
galaxies and radio-loud AGN (e.g. 3C 120, Marshall et al. 2009; 3C 390.3, Gliozzi et al. 2009),
despite a possible contribution from a jet emission in the latter class. However, at the frequencies above
the high-frequency break the soft X-ray PSD of BL Lacs seem to drop more rapidly than it is observed
in Seyfert galaxies.
Hard X-ray (14--150\,keV; Swift/BAT) PSD have been recently constructed by Shimizu \& Mushotzky (2013)
for a sample of 30 AGN, which included two FSRQs (3C 273 and 3C 454.3) and one
BL Lac (Mkn 421). This study covered longer time scales than the soft X-ray study
of Kataoka et al. (2001). A characteristic time scale was detected only in 3C 273 preventing a detailed
cross-band comparison. The PSD of all other AGN in their sample were described with unbroken power-law
functions with a typical slopes, $0<\alpha<2$, when constrained.

The $\gamma$-ray blazar PSD slopes derived in our study are in general consistent with those observed in X-rays in
non-blazar AGN. However, we were
able to derive only the upper limits on the blazar short characteristic time scales. Given that they are consistent
with the low end of $\tau_{\rm S}$ detected in the radio-quiet AGN (Figure~\ref{fig:results}), we cannot discard the possibility that
the $\gamma$-ray $\tau_{\rm S}$ in blazars and the X-ray $\tau_{\rm S}$ in non-blazar AGN are comparable.

However, the OU $\gamma$-ray characteristic time scales detected in PKS 2155-304 and 3C 66A are much longer than those of the BL Lac sources
(and also FSRQs) in the soft X-rays. In addition, closer inspection of few available blazar PSD across different energy bands
suggests that they are rather different. In particular, in PKS 2155-304 the value on $\tau_{\rm S}$ derived in our work is in
disagreement with the X-ray $\tau_{\rm S}$ found in Kataoka et al. (2001).
The PSD slope derived here for 3C 279, $\alpha\simeq1.1$,
is in disagreement with that reported in the soft X-rays by Chatterjee et al. (2008, $\alpha\simeq2.3$). We do not detect a $\gamma$-ray
characteristic time scale in 3C 273, while the soft and hard X-ray detections were reported at the values well within our range of sampled
time scales (McHardy 2008, Shimitzu \& Mushotzky 2013).
These discrepancies imply that either the blazar variability process is energy dependent, or the origins of the X-ray and
$\gamma$-ray blazar variability are different.

The processes intrinsic to the jet physics such as the internal shock model (e.g. Ghisellini 1999) or the
minijets-in-a-jet model (e.g. Begelman et al. 2008, Giannios et al. 2009, Narayan \& Piran 2012) may well dominate
the blazar variability structure leading to the $\gamma$-ray
variability properties that are largely different from those implied by the accretion disk physics and propagation models.
Alternatively, the $\gamma$-ray variability pattern may be
a combination of the baseline stochastic disk/corona variability carried out in a jet, and variability intrinsic to the jet. 
If the propagation model of variability envisaging coupling of perturbations at different radii continues toward the jet, and
the disk X-ray perturbations couple with the X/$\gamma$-ray perturbations intrinsic to the jet, then the resulting 
Fermi/LAT variability structure may differ from that known from the X-ray band. Additional complication may be introduced by
altering the variability time scales due to relativistic effects.

\section{Conclusions}
\label{sec:conclusions}

We have applied the stochastic models of luminosity fluctuations developed in Kelly et al. (2009, 2011) to the
Fermi/LAT light curves of 13 well observed blazars in order to study their time variability properties.
The light curves of 3 blazars were consistent with the OU process which is characterized with a PSD
$\propto 1/f^{\alpha}$ featuring a bend at a characteristic times scale, $\tau_{\rm 0}$, where
the slope, $\alpha$, changes from 0 ($\tau \gg \tau_{\rm 0}$) to 2 ($\tau \ll \tau_{\rm 0}$).
We constrained $\tau_{\rm 0}$ in two BL Lac type blazars,
3C 66A ($\tau_{\rm 0}\simeq25$\,day, or $\simeq17$\,day in the rest frame) and
PKS 2155-304 ($\tau_{\rm 0}\simeq43$\,day, or $\simeq38$\,day in the rest frame).
Thus, the inferred $\gamma$-ray BL Lac characteristic time scales were longer than those
observed in the soft X-ray blazar and Seyfert light curves.
In addition, the low- and high-frequency PSD slopes in the BL Lacs fitted well with the OU process
were flatter than their counterparts in the soft X-rays. These discrepancies indicate either
the energy dependence of the PSD or different origins of the BL Lac X-ray and $\gamma$-ray variability.

In 10 of 13 sources a better agreement with the data was obtained with the mixed OU process characterized with
a PSD featuring two bends and an intermediate part of PSD where $0<\alpha<2$. For these sources we derived
respective limits on the long and short characteristic time scales. We concluded that their underlying PSD has
likely the form of a power-law function over the sampled range of temporal frequencies. The upper limit on the
short characteristic time scale indicates sub-hour variability in 4 FSRQ sources. This finding needs to be
addressed by present and future theoretical models of blazar $\gamma$-ray variability. We constrained the PSD
slopes, $\alpha$, for all sources except 3C 454.3.

Our stochastic approach is particularly well suited for the variability studies of various classes of AGN because
it accounts self-consistently for irregular sampling, measurement errors, red noise leak, and aliasing. Thus, it
provides an alternative to the variability methods based on the PSD construction.

\acknowledgments
This research made use of data obtained with the Fermi Gamma-Ray Observatory.
The authors thank Mitch Begelman and James Chiang for discussions and comments.
Partial support for this work was provided by the Fermi grant NNX11AO45G and by
NASA contract NAS8-03060. M.A.S. was partially supported by the Polish National
Science Center grant No. 2011/03/B/ST9/03281. K.N. was supported by NASA
through Einstein Postdoctoral Fellowship grant No. PF3-140112 awarded by the
Chandra X-ray Center, which is operated by the Smithsonian Astrophysical Observatory
for NASA under contract NAS8-03060.



\begin{table*}
\caption{Properties of the Fermi/LAT blazar light curves.}
\begin{center}
\label{tab:sample}
\begin{tabular}{l r r r r r r r r r r r}
\hline
Name   & $z$ & $\log\tau_{\rm 0}$ &  DIC & $\alpha$ & $\alpha_{\rm NM13}$ & \multicolumn{2}{c}{$\log\tau_{\rm S}$} & \multicolumn{1}{c}{$\log\Delta t$} & \multicolumn{2}{c}{$\log\tau_{\rm L}$} & DIC\\
  &  & (1) & (2) & (3) & (4) & \multicolumn{1}{c}{(5)} & \multicolumn{1}{c}{(6)} & \multicolumn{1}{c}{(7)} & \multicolumn{1}{c}{(8)} & \multicolumn{1}{c}{(9)} & (10)\\
\hline
\multicolumn{12}{l}{FSRQs} \\
\hline
B2 1633+38    & 1.814 & $-0.13^{+0.12}_{-0.12}$ &  1871  & $1.00^{+0.06}_{-0.06}$ & -               &  & $<-1.53$ & $-1.3$ &  & $> 2.27$ & 1663 \\
PKS 1424-41   & 1.522 & $ 0.75^{+0.14}_{-0.12}$ &  1315  & $0.78^{+0.10}_{-0.09}$ & -               &  & $<-0.87$ & $-1.4$ &  & $>2.06$ & 1222\\
B2 1520+31    & 1.487 & $-0.01^{+0.08}_{-0.08}$ &  2458  & $1.05^{+0.08}_{-0.07}$ & -               &  & $<-1.35$ & $-1.3$ &  & $>2.00$ & 2301 \\
PKS 0454-234  & 1.003 & $ 0.25^{+0.11}_{-0.11}$ &  1733  & $0.93^{+0.08}_{-0.07}$ & $0.777\pm0.274$ &  & $<-1.26$ & $-1.3$ &  & $>2.44$  & 1578          \\
3C 454.3      & 0.859 & $ >2.00$                &   254  & $<0.43$                & $0.999\pm0.235$ & $ 0.32^{+1.47}_{-1.22}$ & $<3.21$ & $-0.6$ &  & $>2.53$           & 250 \\
3C 279        & 0.536 & $-0.01^{+0.09}_{-0.09}$ & 2425  & $1.03^{+0.06}_{-0.06}$ & $1.078\pm0.246$ &  & $<-1.53$ & $-1.3$ &  & $>2.39$ & 2169 \\
PKS 1510-089  & 0.360 & $ 1.03^{+0.11}_{-0.10}$ & 1386   & $0.48^{+0.19}_{-0.20}$ & $1.101\pm0.298$ &  & $<-0.16$ & $-0.6$ & $2.17^{+1.52}_{-0.45}$  & $>1.64$           & 1350\\
3C 273        & 0.158 & $ 0.38^{+0.10}_{-0.10}$ &  1509   & $0.84^{+0.11}_{-0.09}$ & $1.301\pm0.265$ &  & $<-1.03$ & $-1.6$ &  & $>2.06$       & 1408    \\
\hline
\multicolumn{12}{l}{BL Lacs} \\
\hline
3C 66A        & 0.444 & $1.39^{+0.43}_{-0.29}$ & 448    & $0.81^{+0.50}_{-0.44}$ & $0.604\pm0.438$ & $-0.03^{+0.76}_{-1.19}$ & $<1.03$ & $-0.7$ &  & $>1.87$          & 445   \\
PKS 0716+714  & 0.300 & $0.17^{+0.10}_{-0.10}$ & 1941   & $1.05^{+0.20}_{-0.16}$ & -               & $-1.10^{+0.44}_{-0.73}$ & $<-0.53$ & $-1.3$ &  & $>1.82$  &  1876\\
PKS 2155-304  & 0.116 & $1.63^{+2.79}_{-0.35}$ & 477   & $0.64^{+0.79}_{-0.50}$ & $0.577\pm0.332$ &  & $<1.61$ & $-0.7$ &  & $>1.70$   & 468         \\
BL Lac        & 0.069 & $0.47^{+0.15}_{-0.15}$ & 865   & $0.93^{+0.18}_{-0.14}$ & $0.412\pm0.469$ & $-1.16^{+0.52}_{-0.72}$ & $<-0.46$ & $-1.3$ &  & $>2.11$        & 811    \\
Mkn 421       & 0.030 & $0.82^{+0.15}_{-0.13}$ & 1351  & $0.94^{+0.13}_{-0.13}$ & $0.384\pm0.205$ &  & $<-0.51$ & $-0.8$ &  & $>1.78$         & 1306  \\

\hline

\end{tabular}\\
\end{center}
{\bf Notes:} All characteristic time scales are reported in days and in the observer's frame.
(1) Characteristic time scale of the OU process, $\tau_{\rm 0}$, in days.
(2/10) Deviance Information Criterion, DIC, for the OU/sup-OU model; models with smaller DIC are preferred.
(3) Slope of the ${\rm PSD} \propto 1/f^{\alpha}$ between the short and long characteristic time scales in the sup-OU model (90\% confidence region).
(4) Slope of the ${\rm PSD} \propto 1/f^{\alpha}$ derived by NM13 through the Fermi/LAT PSD construction.
(5) Short characteristic time scale in the sup-OU model, $\tau_{\rm S}$, in days (90\% confidence region).
(6) Upper limits on the short characteristic time scale in the sup-OU model (99\% confidence region).
(7) The smallest time spacing in the light curve, $\Delta t$, in days.
(8) Long characteristic time scale in the sup-OU model, $\tau_{\rm L}$, in days (90\% confidence region).
(9) Lower limit on the long characteristic time scale in the sup-OU model (99\% confidence region).
\end{table*}


\clearpage

\end{document}